\documentclass[aps, amssymb, amsmath, reprint]{revtex4-1} 

\usepackage{graphicx}  
\usepackage{dcolumn}   
\usepackage{physics}
\usepackage{color}
\usepackage{appendix}
\usepackage{ulem}
\usepackage{soul}
\definecolor{orange}{rgb}{0.8, 0.3, 0}

\definecolor{blueviolet}{rgb}{0.2, 0.2, 0.6}
\usepackage[pdftex, 
bookmarks=true, 
colorlinks=true,
allcolors=blueviolet,
pdfstartview={FitH}, 
]{hyperref}

\newcommand{\gens}{g_\text{ens}}

\newcommand{\Er}{\text{Er}^{3+}}

\newcommand{\trep}{t_\text{rep}}
\newcommand{\tpump}{t_\text{pump}}
\newcommand{\Tsl}{T_1^\text{sl}}
\newcommand{\ginh}{\Gamma_\text{inh} }

\begin{document}

\title{Phonon-bottlenecked spin relaxation of Er$^{3+}$:CaWO$_4$ at milliKelvin temperatures }

\author{S.~Rajendran$^{1}$}
\author{B.~Mistri$^{1,2,3}$}
\author{P.~K.~Sharma$^{1}$}
\author{S.~E.~Kubatkin$^4$}
\author{A.~V.~Danilov$^4$}
\author{S.~Dhomkar$^{2,3}$}
\author{S.~E.~de~Graaf$^{5}$}
\author{V.~Ranjan$^{1}$}
\email{vranjan@tifrh.res.in}

\affiliation{$^1$ Tata Institute of Fundamental Research, Hyderabad 500046, India}
\affiliation{$^2$ Department of Physics, Indian Institute of Technology Madras, Chennai 600036, India}
\affiliation{$^3$ Center for Quantum Information, Communication and Computing, Indian Institute of Technology Madras, Chennai 600036, India}
\affiliation{$^4$Department of Microtechnology and Nanoscience MC2, Chalmers University of Technology, SE-41296 Goteborg, Sweden}
\affiliation{$^5$National Physical Laboratory, Teddington TW11 0LW, United Kingdom }

\begin{abstract} 
We study spin-lattice relaxation times of electron spins in $\Er:\text{CaWO}_4$ at millikelvin temperature, detected via their coupling to a low-mode volume superconducting resonator. At large magnetic field supporting strong phonon-emission rates, we observe a noticeable increase in relaxation times with increasing spin-excitations, which exhibit a unique  $[\tanh (\hbar \omega_0/k_\text{B} T)]^2$ temperature dependence. These observations are typical of a phonon-bottlenecked spin relaxation, and have implications for quantum technologies that exploit rare-earth spin ensembles as coherent resources. 
\end{abstract}

\maketitle
Rare-earth ion doped crystals have emerged as promising candidates for quantum information processing as they enable microwave-to-optical transduction and the realization of optical quantum memories~\cite{lvovsky_optical_2009, hedges_efficient_2010, sinclair_spectral_2014, businger_optical_2020}. Among these, $\Er$ has garnered considerable interest due to its optical transition within the telecom C-band, making it well-suited for low-loss fiber-based quantum networks~\cite{sangouard_quantum_2011,lauritzen_telecommunication-wavelength_2010, xie_scalable_2025, rochman_microwave--optical_2023}. Furthermore, GHz-range operation under sub-tesla magnetic fields positions this material as a viable platform for microwave quantum memories and coherent interfaces with superconducting qubit architectures~\cite{probst_anisotropic_2013, dantec_twenty-threemillisecond_2021, rancic_electron-spin_2022, ranjan_spin-echo_2022, alexander_coherent_2022, gupta_robust_2023}. Practical implementation, however, requires cooling to millikelvin temperatures, both to suppress thermal noise and to mitigate spectral diffusion, the latter being especially critical in heavily doped samples where high concentrations are necessary for efficient photon absorption~\cite{afzelius_proposal_2013, morton_storing_2018}. Under such cryogenic conditions, spin coherence time reaching 23~ms~\cite{dantec_twenty-threemillisecond_2021} has been reported in erbium doped CaWO$_4$ crystals establishing them as a leading host material.

The upper limit to spin coherence times is set by the spin to bath relaxation time $T_1$. For $\Er:\text{CaWO}_4$, the most notable study on $T_1$ relaxation date back to the 1960s in which direct phonon and two-phonon Orbach processes were found as the dominant relaxation mechanisms in the temperature range $1-20$~K~\cite{antipin_paramagnetic_1968}. Indeed, for Kramers ions, lattice phonons strongly couple to spins via thermal modulation of the crystal field and spin-orbit coupling. Consequently, fast relaxation rates can exist even at mK temperatures despite freezing of resonant phonons~\cite{abragam_electron_2012}. This is in contrast to other popular systems such as group V defects in diamond or silicon, where $T_1$ is found to approach several hours under similar measurement conditions~\cite{astner_solid-state_2018, bienfait_controlling_2016}. Especially at mK temperatures, additional complications to direct phonon processes can arise since there are not enough resonant phonons available to transport the excitations from spins to the thermal bath. This so-called phonon bottleneck (PB) can have an unfavorable effect through a buildup of excitations in the resonant lattice modes, unintended re-excitation of spins, and subsequently a sudden release of spin energy via a phonon-avalanche~\cite{brya_paramagnetic_1965}. These phenomena have parallels to an inverted gain medium for photons and their stimulated emission in two-level lasers and masers. However, a controlled PB can be turned into a resource, as recently illustrated through $T_1$ enhancement of two-level system defects (TLSs) coupled to an engineered phononic bandgap~\cite{chen_phonon_2024}. PB effect was first predicted by van Vleck~\cite{van_vleck_paramagnetic_1941}, and since then measured in variety of magnetic systems~\cite{scott_spin-lattice_1962, ruby_paramagnetic_1962,wagner_fractal_1985,chiorescu_butterfly_2000, faber_hot_2024}. It has recently regained attention in the context of quantum technologies~\cite{budoyo_electron_2018, soriano_exploring_2025, chiossi_optical_2026}. Nonetheless, spin relaxation studies in highly doped rare-earth crystals at millikelvin temperatures remain challenging due to complications such as spin-spin interactions and cross-relaxation from other impurity spins or hyperfine transitions.

In this Letter, we report the observation of phonon-bottleneck effects at dilution refrigerator temperatures and nW power excitation regimes in electron spin ensembles of $\Er:\text{CaWO}_4$. The PB leads to enhanced spin to bath relaxation times of the spins, and is most noticeably seen in the unique temperature dependence of spin polarization decay exponentials following $[\tanh (\hbar \omega_0/k_\text{B} T)]^2$, and is supported by spin excitation number and field dependencies. In particular, we find PB to be more pronounced with a narrower spin inhomogeneous linewidth. The magnetic field dependence of the spontaneous phonon emission is also studied, verifying weaker spin-phonon couplings at lower magnetic fields in Kramers ions.

Our measurements are performed on a nominally 50~ppm doped erbium doped CaWO$_4$ crystal purchased from SurfaceNet Germany. The $\Er$ ions substitute Ca$^{2+}$ sites and form an effective electron spin $S=1/2$ system at low temperatures. Though erbium has several isotopes with natural abundance, these can be categorized by distinct nuclear spins $I=0$ (78\%) and $I=7/2$ (22\%). Owing to highly anisotropic g-factors, though diagonalizable along crystallographic axes (g$_{a}$ = g$_{b}$ = 8.38, g$_{c}$ = 1.24)~\cite{enrique_optical_1971, abragam_electron_2012}, the magnetic field position of spin resonances change with the relative orientation of the external magnetic field. This is also accompanied by changes in the inhomogeneous spin-linewidth from coupling to randomly distributed static charge defects~\cite{mims_broadening_1966, dantec_twenty-threemillisecond_2021}. This, in turn, allows for different effective spin densities for different field orientations. In addition, the hyperfine transitions provide a naturally lower concentration of spin species in the same crystal environment. 

\begin{figure}[t!]
    \centering
\includegraphics[width=\columnwidth]{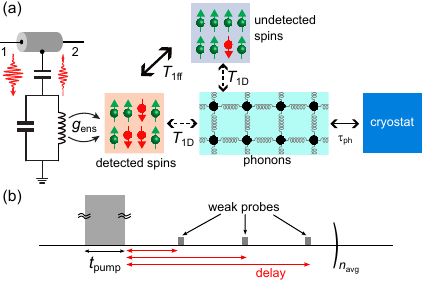} 
\caption{Measurement setup. (a) Schematics of excitation and subsequent relaxation of electron spins coupled to a superconducting resonator with a bandwidth $\kappa \ll \Gamma_\text{inh}$, the inhomogeneous spin linewidth. For large spin concentrations and at mK temperatures, excited spins are most likely to exit the detection volume via spin flip-flops ($T_\text{1ff}$), and only weakly via direct phonon emission ($T_\text{1D}$). $\tau_\text{ph}$ is the average time for phonons to equilibrate with the cryostat. (b) Pulse sequences used for extracting spin relaxation by measuring resonator response following a long and strong saturation pump.  }
    \label{fig1:Schematics}
\end{figure}

The spin crystal is attached using vacuum grease on top of a NbN superconducting resonator~\cite{de_graaf_direct_2017, mahashabde_fast_2020} fabricated on saphhire substrate, operating at a frequency $\omega_0/2\pi = 4.44$~GHz and capacitively coupled to the feedline with a loss rate of $\kappa_c/2\pi = 2.5 \pm 0.1~$MHz. The sample is mounted inside a dilution refrigerator with the base temperature of 20~mK. We apply magnetic field $B_\text{mag}$ in the $ac$ plane of the crystal at an angle $\theta$ with the $c$-axis. The field is kept parallel to the superconducting thin film to avoid resonator losses from magnetic vortices. All spin relaxation measurements shown in the main text are performed on the $I=0$ transition, with control pulses generated using an arbitrary waveform generator, and signal quadratures acquired through homodyne demodulation at GSa/s. More details of the superconducting resonator and spin system are presented in our previous works \cite{ranjan_spin-echo_2022} and the supplementary information.


\begin{figure}[t]
    \centering
\includegraphics[width=\columnwidth]{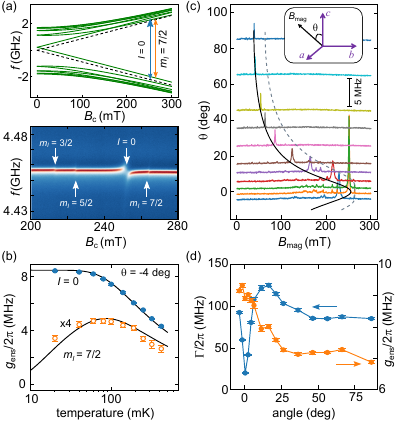} 
\caption{Spectroscopy of $\Er$ spins in CaWO$_4$. (a) Top: Energy eigenvalues of two isotopes of $\Er$ with nuclear spin $I=0$ (dashed lines) and $7/2$ (solid lines).  $m_I$ is the nuclear-spin projection on the B-field axis. Bottom: Color plot of transmission power $|S_{21}|^2$ vs magnetic field at 80~mK. The field is aligned with the crystal $c$-axis. (b) Temperature dependence of $\gens$ (symbols: measurements, lines: theory) for the two transitions marked by double arrows in the top panel of (a). (c) Angular dependence of the cw spin spectra at 20~mK. Largest peaks correspond to the erbium $I=0$ transition. Angle $\theta$ is with respect to $c$-axis in the $ac$ plane. Solid and dashed lines are calculated using the electron spin Hamiltonian for $I=0$ and $m_I=7/2$ transitions, respectively. Data have a vertical offset, proportional to the measured angle. (d) Angular dependence of extracted $\gens$ (right-axis) and $\Gamma$ (left axis) for the $I=0$ transition at 20~mK.}
    \label{fig:spectra}
\end{figure}

Calculated energy eigenvalues of two Er species are plotted in top panel of Fig.~\ref{fig:spectra}(a), for two Er isotopes, with the magnetic field along the $c$-axis ($\theta = 0$). In the limit of a large magnetic field, only nuclear spin preserving transitions are allowed. Two such resonant transitions are marked by double arrows in the calculated spectra and their field position identified in the continuous wave (cw) transmission measurements in Fig.~\ref{fig:spectra}(d) acquired at $T=80$~mK. $m_I$ is the nuclear-spin projection on the $B$-field axis. The relatively large response observed for the $I=0$ transition is consistent with its higher abundance. 

We perform temperature dependence of cw measurements and extract the spin-resonator ensemble coupling strength $\gens$ by fitting the change in total resonator loss to $\Delta\kappa = \gens^2 \ginh/\left[ (\omega - \omega_s)^2 + (\ginh/2)^2 \right]$, where $\gens = g_0 \sqrt{N}$, with $N$ the total number of spins and $g_0$ the single spin-resonator coupling strength. We note that $g_0$ in our case is extremely inhomogeneous owing to the planar resonator (inductor width $1~\mu$m) and bulk crystal doping. The extracted $\gens$ for different temperatures and two transitions are plotted in Fig.~\ref{fig:spectra}(b), and found to be consistent with the theoretical predictions. The non-monotonic dependence of $\gens$ for $m_I = 7/2$, in particular, helps deduce a spin temperature of $\approx 50~$mk at the base temperature of the cryostat.

Extracted $\kappa$ versus magnetic field magnitude $B_\text{mag}$ is plotted for different angles $\theta$ in Fig.~\ref{fig:spectra}(c). In accordance with the g-factor anisotropy, the resonance magnetic field position $B_0$ for the $I=0$ transition changes from 255~mT at $\theta =0$ to 38.5~mT at $\theta = 90$ degrees. We further relate the height and width of the resonance to $\gens$ and $\ginh$, respectively, with temperature and angular dependence shown in Fig.~\ref{fig:spectra}. That $\gens$ reduces by roughly a factor of $1/\sqrt{2}$ near $\theta = 90$~deg is expected from the perpendicular component of the g-tensor. The corresponding $\ginh$ [left axis in Fig.~\ref{fig:spectra}(d)] shows strong variations from $28 \pm 3$~MHz to $125 \pm 10$ around $-10 <\theta < 10$, highlighting the symmetry c-axis along which the inhomogeneity of the electric field from charge defects is supposed to be maximally suppressed~\cite{mims_broadening_1966}. The residual line width at $\theta =0$ can be mainly attributed to the remaining misalignment of the field direction with the crystal c-axis.

We first perform echo inversion recovery sequences to study spin relaxation processes. However, strong spin flip-flop rates in our heavily doped sample cause spin excitations to leave our small detection volume before they can relax to the thermal bath [Fig.~\ref{fig1:Schematics}(a)]. Their typical $[\text{sech}(\hbar \omega_s/k_\text{B}T)]^2$ temperature dependence of relaxation rates~\cite{bottger_optical_2006,car_optical_2019,rancic_electron-spin_2022}  is shown in the supplementary. To access slower spin-lattice relaxation processes, we instead weakly probe (readout power $\sim 1~$pW, duration 160~$~\mu$s) the response of the resonator after a relatively strong saturation pump (power 2.5~nW) incident on the input of the feedline [Fig.~\ref{fig1:Schematics}(c)]. These pulses are specifically chosen to be longer than the spectral diffusion time to allow for establishing spin depolarization over a larger mode volume, and whose re-population over time appears in the resonator loss rate $\kappa$, and hence the transmitted field across the feedline~\cite{amsuss_cavity_2011, ranjan_probing_2013, probst_anisotropic_2013}. 

\begin{figure}
    \centering
\includegraphics[width=\columnwidth]{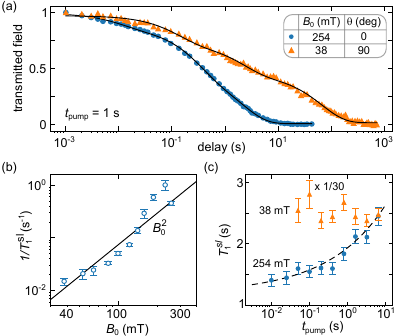}
\caption{Spin relaxation via direct phonon processes. (a) Normalized transmitted field across the feedline (symbols: measurement, lines: triple-exponential fits) following a long saturation pulse of duration 1~s at two magnetic fields. Data for $\theta=0$ (90) is trace averaged 40 (5) times. (b) Magnetic field dependence of the slow relaxation time constant $T_1^{sl}$. Solid line represents $B_0^2$ dependence. Note the semi-log and log-log plot in panels (a,b), respectively. (c) Dependence of $T_1^{sl}$ on the pump duration for two magnetic fields. Error bars represent the standard deviations associated with the least square fits of $\Tsl$. }
    \label{fig:T1D}
\end{figure}

The transmitted signal measured after a pump pulse of duration $\tpump = 1$~s at two angles is plotted in Fig.~\ref{fig:T1D}(a). We observe complex signal decays for which a triple-exponential fit works best. For $\theta =0$, we obtain (fast, medium, slow) time constants of  $T_{1}^\text{f} = 10.7 \pm 1~\text{ms}, T_{1}^\text{m}= 298 \pm 30~\text{ms}, \Tsl = 2.2 \pm 0.2~\text{s}$ with amplitude weights of $13\%, 41\%,$ and $ 46\% $, respectively. The relative weights are found to change with the pump parameters, although no correlation is observed between the time constants and the respective weights. In the following, we focus our discussion solely on the slowest time constant $\Tsl$, as we aim to understand the most limiting process that determines the energy transfer from spins to the thermal bath. We, however, note that $T_{1}^\text{m}$ shows a behavior similar to $\Tsl$ across experimental conditions, while $T_1^\text{f}$ is found to arise from non-spin origins (see the supplementary). 

We make an independent estimate of $\Tsl$ by varying the repetition time of standard echo-inversion recovery sequences (supplementary). A short repetition time does not allow all spins to return to thermal equilibrium and thus results in an accumulation of spin excitation and an effective higher spin temperature. This, in turn, leads to enhanced spin flip-flop rates reflected in their measured spectral diffusion times. Using this method, we obtain a similar $\Tsl (\theta = 0) \sim 1.5$~s, corroborating $\Tsl$ to be the spin-to-bath relaxation time. 

Spin relaxation measurements are then performed at other angles corresponding to different resonance magnetic field positions $B_0$. For $\theta = 90$~degrees and $B_0 = 38~$mT, we clearly see the signal decay extends by almost two orders of magnitude compared to $\theta =0$. For $\theta =0$ we obtain a $\Tsl = 67 \pm 7$~s [Fig.~\ref{fig:T1D}(a)]. The extracted $\Tsl$ versus the entire magnetic field range is plotted in Fig.~\ref{fig:T1D}(b). The observed increase in $T^\text{sl}_1$ with the decrease in magnetic field can be explained by the direct phonon process for Kramers ions, which at a constant resonance frequency~\cite{larson_spin-lattice_1966} is given as

\begin{align} \label{eq:T1d}
1/T_\text{1D} = \alpha_\text{D}\omega_s^3 B_0^2  \coth\left(\frac{\hbar \omega_0}{2 k_B T} \right),
\end{align}
where $\alpha_\text{D}$ contains an anisotropic transition matrix element, fundamental constants, and crystal properties. At the lowest temperature of 20~mK, the observed values $\Tsl$ correspond to the spontaneous phonon emission process, with an approximate quadratic dependence $B_0^2$ [plotted as a solid line on a logarithmic scale in Fig.~\ref{fig:T1D}(b)]. Some deviations around 200~mT can be attributed to the angle anisotropy of the spin-lattice coupling~\cite{larson_spin-lattice_1966, antipin_paramagnetic_1968}. The second-long time constants at 20~mK observed here are consistent with those previously reported for $\Er$ spins hosted in YSO~\cite{probst_anisotropic_2013}, CaWO$_4$~\cite{dantec_twenty-threemillisecond_2021,billaud_electron_2025} and YO~\cite{gupta_robust_2023}. As there is no visible saturation in $\Tsl$ due to Purcell spin relaxation~\cite{purcell_spontaneous_1946, bienfait_controlling_2016, eichler_electron_2017}, we can estimate, from single spin perspective, an upper bound of the coupling strength $g_0/2\pi < 20~$Hz and Rabi frequency $\Omega_\text{R}=2g_0\sqrt{n}/2\pi \approx 400~$kHz for the pump power used. 

\begin{figure}
    \centering
\includegraphics[width=\columnwidth]{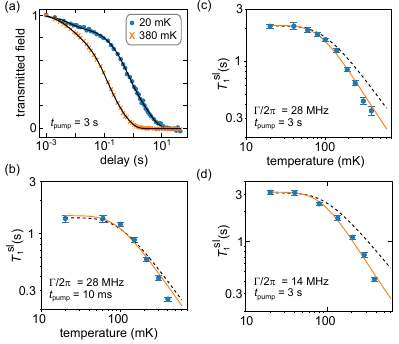}
\caption{ Phonon bottlenecked spin to bath relaxation. (a) Transmitted field amplitude at two temperatures, symbols: measurement, lines: triple-exponential fits. (b-d) Temperature dependence of the slowest time constants for different experimental conditions, the inhomogeneous spin linewidth $\ginh$ and pump duration $\tpump$. Dashed and solid lines are fits to the direct phonon process alone and the phonon-bottleneck mediated spin relaxation process, respectively. Error bars represent the standard deviations associated with the least square fits of $\Tsl$. }
    \label{fig:pb}
\end{figure}

To gain further insight into the spin-lattice relaxation, we measured the temperature dependence at $\theta = 0$. As shown in Fig.~\ref{fig:pb}(a), the transmitted field maintains a similar three-exponential decay across the temperature range. The extracted $\Tsl$ are plotted in Fig.~\ref{fig:pb}(b-d) for two pump durations and two  inhomogeneous linewidths. The linewidth corresponding to $\ginh/2\pi=28\pm 3~(14 \pm 2)~$MHz was obtained at the same angle $\theta =0$ (lab frame) and magnetic field $B_0 = 254 \pm 0.3$ mT but in a separate fridge cooldown, and caused by natural misalignment with the $c$-axis. We first observe that even at the lowest temperature of 20~mK, $\Tsl$ values are different for different conditions, which is not expected from direct phonon processes. In contrast, no dependence of $\Tsl$ on pump duration is observed at 38~mT [Fig.~\ref{fig:T1D}(c)]. To understand the extent of these discrepancies, fits are made to the direct phonon process, and are plotted as dashed lines in Fig.~\ref{fig:pb}. Although the fit to the data is somewhat good for the shorter pump duration $\tpump=10$~ms [Fig.~\ref{fig:pb}(b)], significant deviations are present for panels (c,d).

The aforementioned observations are consistent with a phonon-bottleneck, which is expected to produce a highly non-linear relaxation response. The effect can be analytically quantified in the weak-excitation or linear approximation regime~\cite{scott_spin-lattice_1962}, predicting a modified spin-lattice relaxation time $T_\text{1D} \rightarrow T_\text{1D} + T_{1b} $, with
\begin{align} \label{eq:Tb}
1/T_{1b} = \frac{1}{\tau_\text{ph} }\frac{3\omega_0^2 \Gamma_\text{inh} }{2 \pi^2 c v^3} \left [ \coth\left(\frac{\hbar \omega_0}{2 k_\text{B} T} \right) \right ]^2.
\end{align}
Here $c$ is the spin concentration, $v$ the speed of sound and $\tau_\text{ph}$ the thermalization time of the crystal phonons with the bath. The most striking observation for PB is in its distinct temperature dependence. To verify this, we perform a fit to $\Tsl = T_\text{1D}^0 \tanh (\frac{\hbar \omega_0}{2 k_\text{B} T}) + T_\text{1b}^0 \left [ \tanh (\frac{\hbar \omega_0}{2 k_\text{B} T}) \right ]^2$, where we phenomenologically absorb the non-linear strength of PB through $T_\text{1b}^0$. Plotted as solid line, we are able to obtain good fits to the data using a common $T_\text{1D}^0 = 1.2 \pm 0.1$~s, representing the bare time constant in absence of PB, and variable $T_\text{1b}^0 = 0.25 \pm 0.02$~s, $0.9 \pm 0.1$~s and $1.95 \pm 0.2$~s for figure panels (b), (c) and (d), respectively. In particular for panel (d), the bottleneck correction $T_\text{1b}^0$ exceeds the bare $T_\text{1D}^0$ by $160\%$.

It can be seen from Eq.~\ref{eq:Tb} that the PB can be suppressed with a lower spin density $c/\ginh$. This is readily noticed when comparing data for two $\ginh$ at the same $\tpump$, In fact, the extracted doubling of $T_\text{1b}^0$ is consistent with halved $\ginh$. Larger $\ginh$ naturally leads to fewer spin excitations for our narrow-bandwidth pulse excitation. A similar qualitative effect is achieved with shorter pump duration, e.g. $\tpump = 10$~ms in panel (b), resulting in a weaker PB correction $T_\text{1b}^0= 0.25~$s. The lack of response in $\Tsl$ to changing pump duration at $38~$mT in Fig.~\ref{fig:T1D}(c) is also consistent with the theory since for this case $T_\text{1D} \gg T_{1b}$. 

Finally, we theoretically estimate the strength of PB using the prefactor in Eq.~\ref{eq:Tb}. Taking  $v \approx 3000~$~m/s~\cite{gluyas_elastic_1973}, $c = 5 \times 10^{17}~\text{cm}^{-3}$ and $\tau_\text{ph} \approx d/v$, where $d = 0.2~$mm is thickness of the crystal, we find $T_\text{1b}^0 \approx 0.1~$s. The discrepancy from the experimental value possibly arises from the finite bandwidth of the spin excitation $\Omega_\text{R}\ll \ginh$, but also because $\tau_\text{ph}$ can be much longer than the estimated value of $d/v$ in our setup in which the crystal is simply glued on top of the resonator with vacuum grease. It is interesting to compare our experiments with the first observation of PB in electron spin ensembles ($1\%$ Pr in La$MN$) performed at $T = 1.4$~K with 2~W saturation power~\cite{scott_spin-lattice_1962}, whereas we are able to detect the same with nW power. This highlights the operation at mK temperatures, allowing much higher spin polarization, and larger coupling strength $g_0$ possible with small mode volume superconducting resonators~\cite{probst_inductive-detection_2017, ranjan_electron_2020}.  

In summary, we have measured contributions of direct phonon emission and phonon-bottleneck toward spin relaxation across varying experimental conditions in a 50 ppm $\Er$ doped CaWO$_4$ crystal. We are able to explain the strong angular, magnetic field, and temperature dependencies using simple theoretical descriptions, though further studies are needed to fully understand the relative weights and time constants of the multi-exponential decays~\cite{antipin_paramagnetic_1968, zhang_optical_2024, becker_zeeman_2026}. The direct-phonon relaxation rate carries no concentration dependence, thus a non-uniform doping in the crystal can be easily ruled out. In fact, it has been shown that phonon-bottlenecked lattices can cause phonon-avalanches due to stimulated phonon-emission, resulting in a fast and strongly non-linear thermalization between spins and phonons before the usual slow bottlenecked relaxation starts to release energy to the bath~\cite{brya_paramagnetic_1965}. Understanding these relaxation mechanisms will be vital to mitigating associated unfavorable effects, and thus to the potential use of $\Er$ spin ensembles as coherent resources for quantum technologies at mk temperatures, as protocols involving high-power control pulses, such as in quantum memory state retrievals~\cite{osullivan_random-access_2022, ranjan_spin-echo_2022} or microwave to optical transduction~\cite{rochman_microwave--optical_2023} will necessitate quick thermalization of the phonons with the bath.

The authors thank Patrice Bertet for providing useful comments on the manuscript. This work is supported by Department of Atomic Energy, Government of India, under Project Identification No. RTI 40007, and the Department of science and technology (DST) India for the SERB startup grant SRG/2023/001999. S.D. acknowledges the Indian Institute of Technology Madras, India, and the Science and Engineering Research Board (SERB Grant No. SRG/2023/000322), India, for financial support through start-up funding. S.D. and B.M. also acknowledge the support provided by the Mphasis F1 Foundation to the Center for Quantum Information, Communication, and Computing (CQuICC). The resonator sample was fabricated at Myfab Chalmers. SDG acknowledges support from the UK Department for Science Innovation and Technology through the UK National Quantum Technologies Programme. The data that support the findings of this study are available from the corresponding author upon reasonable request.
\newpage
\newpage
\renewcommand{\thefigure}{S\arabic{figure}}
\setcounter{figure}{0}
\renewcommand{\theequation}{S\arabic{equation}}
\setcounter{equation}{0}
\renewcommand{\thesection}{S\arabic{section}}
\setcounter{section}{0}
\section*{Supplementary Information}

\section{Measurement setup}
\begin{figure}[htbp]
    \centering
\includegraphics[width=\columnwidth]{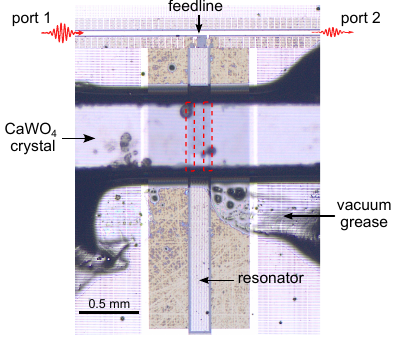} 
\caption{Resonator-spin system. The crystal is attached on top of the resonator using vacuum grease. Marked by dashed boxes, the active area of the resonator over which spins in crystal couple to the resonator is only around 250~$\mu$m long inductor. The signal is sent on port 1 and the transmitted signal measured through port 2 of the feedline. }
    \label{supplfig:resonator}
\end{figure}

The superconducting resonator is patterned from a 50~nm thick NbN film deposited on a sapphire substrate. An optical image showing the placement of the CawO$_4$ crystal is shown in Fig.~\ref{supplfig:resonator}. The resonator capacitors are fractal in shape, except in the area on which the crystal is glued (marked as dashed boxes). Here, the coupling inductors are straight and have a width of $1~\mu$m. Electromagnetic simulations of similar resonators are discussed in the supplementary information of Ref.~\cite{ranjan_spin-echo_2022}. 

\begin{figure}[htbp]
    \centering
\includegraphics[width=\columnwidth]{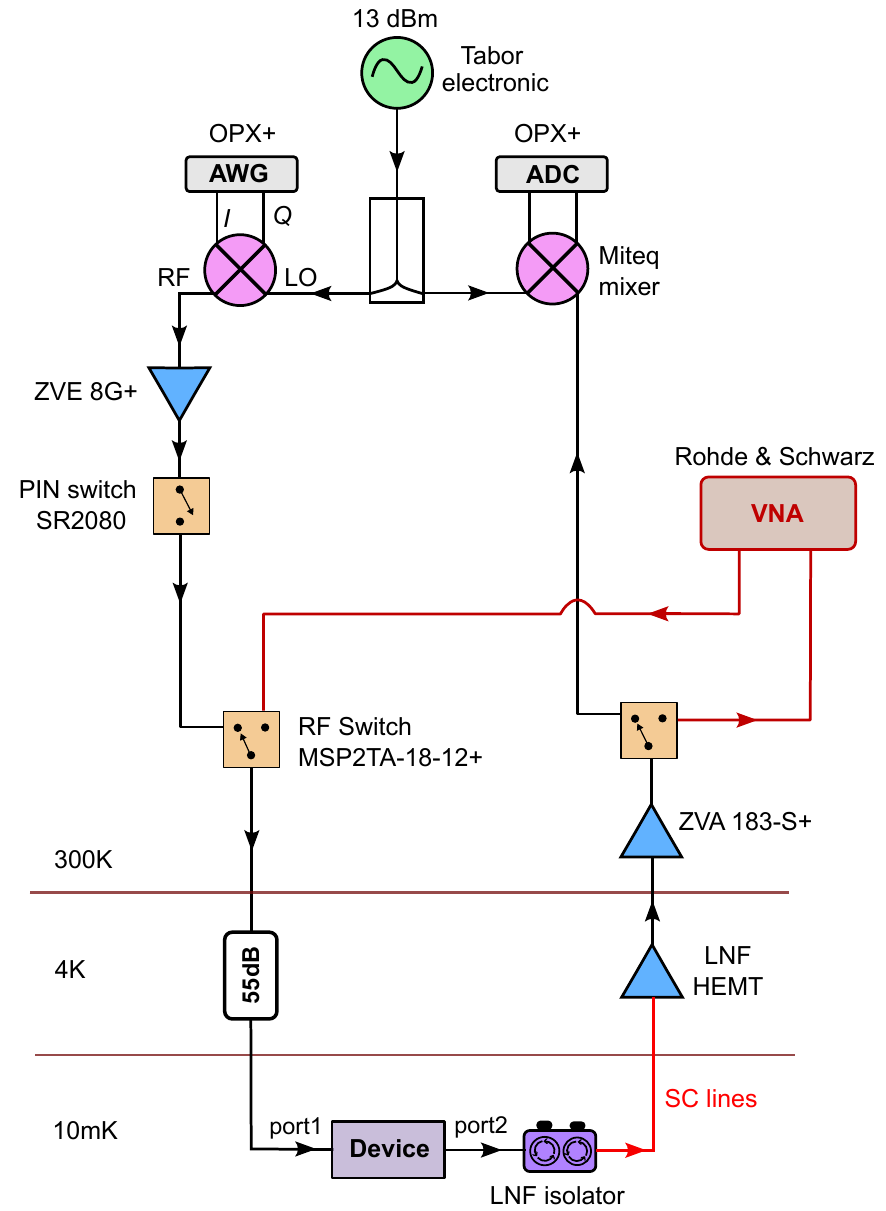} 
\caption{Configurations for pulsed and cw measurements. The sample is mounted at the lowest plate of a dilution refrigerator. The input line is attenuated to prevent thermal noise from reaching the sample. The HEMT mounted at 4K stage with a noise temperature of 1.2K determines the signal to noise ratio. Superconducting coaxial lines are installed between the device output and HEMT input to suppress the insertion losses. Components at other temperature plates have been consolidated for clarity.}
    \label{supplfig:lines}
\end{figure}

Sample is mounted at the base temperature of a Bluefors dilution refrigerator, equipped with a vector magnet. The field axis, perpendicular to the resonator surface, however, is not used. A minimalist setup of measurement lines is shown in Fig.~\ref{supplfig:lines}. Continuous wave measurements are done using a VNA. Pulses are generated by mixing a low frequency signal from an arbitrary waveform generator (Quantum Machines' OPX) with the carrier frequency from a local oscillator (LO). The RF signal after the mixer is amplified using a high-saturation power amplifier to offset the attenuation in the input lines. The thermal noise and any leakage power from the setup are prevented from reaching the sample using a PIN diode switch providing 60 dB isolation. The output signal is first amplified at the 4~K stage of the fridge using a HEMT that has a nominal noise temperature of 1.2~K. Output signals from pulsed measurement are downconverted using another mixer proceeding to homodyne detection using the acquisition module of the OPX.

Two measurement schemes have been employed to extract spin relaxation time constants in this work. An inversion recovery sequence starts with a $\pi$ pulse (duration = $4~\mu$s) to invert the spin population, whose repopulation after a certain delay is probed using a two-pulse (duration = $2~\mu$s) Hahn echo. Echoes at different delays are acquired at a repetition time of $\trep$. Secondly, a pump-probe sequence is employed. Following a saturation pump (power $\sim$ 2.5~nW, duration = $\tpump$) applied at spin-resonance, the resonator is weakly probed (power $\sim$ 1~pW, duration = 160~$\mu$s) successively at several delays~\cite{amsuss_cavity_2011, ranjan_probing_2013, probst_anisotropic_2013} via transmitted field across the feedline. Since the resonator bandwidth is much smaller than the spin linewidth ($\kappa \ll \Gamma_\text{inh}$), control pulses excite only a fraction of the spin population. Both schemes are thus somewhat similar to probing relaxation dynamics using optical spectral hole burning ~\cite{bottger_optical_2006}.

\section{Spin Spectroscopy}
\begin{figure}[htbp]
    \centering
\includegraphics[width=\columnwidth]{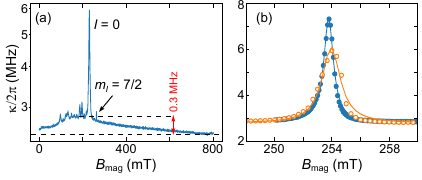} 
\caption{Continuous wave spectroscopy. (a) Extracted $\kappa$ over a larger magnetic field range at $\theta = -4$~deg, showing nuclear spin free $\Er$ as the dominant spin species. A broad background can also be seen, which is typical of measurements using superconducting circuits on sapphire~\cite{de_graaf_direct_2017}. Note the semi-log plot. (b) Extracted $\kappa$ (symbols: measurement. lines: fits) at $\theta = 0$ acquired in two different cooldowns, corresponding to different inhomogeneous spin linewidths. }
    \label{supplfig:Spectra}
\end{figure}

We performed an extended magnetic field sweep up to 800~mT to track spin species in the device. As shown in Fig.~\ref{supplfig:Spectra}, no transitions are detected to have an intensity comparable to that of the $I=0$ erbium peak, the transition studied here. The next strongest transition observed is in fact from the hyperfine levels of erbium, with the closest ones e.g. $m_I=7/2$ having $< 40$ times spin concentration [Fig.~2(b) in the main text]. Therefore, we do not expect significant cross-relaxation ($<2\%$) in our measurements~\cite{larson_spin-lattice_1966} from the hyperfine manifold. We, however, also observe a broad background with a spin linewidth larger than 100~mT, even existing till zero field. Such spins have been previously reported to reside on sapphire surfaces~\cite{de_graaf_direct_2017} on which the resonator is patterned.

Our measurements were performed across two cryostat cooldowns that fortunately yielded different spin inhomogeneous linewidths at $\theta =0$ (angle in the laboratory frame) due to natural misalignment of the crystal c-axis with the field axes. For reliable extraction of $\ginh$, cw spectroscopy measurements are performed at a higher temperature of 300~(400) ~ mK with fits yielding $\gens/2\pi = 4.8\pm 0.5~(4\pm0.5)~$MHz, and $\ginh/2\pi = 28 \pm 3~(14 \pm 3)~$MHz [Fig.~\ref{supplfig:Spectra}(b)]. The values outside and inside brackets are for the two cooldowns respectively.

\section{Spin diffusion}
\begin{figure}[t!]
    \centering
\includegraphics[width=\columnwidth]{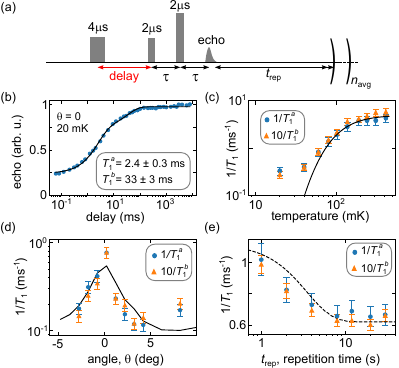} 
\caption{Spin relaxation via spin flip-flops. (a) A standard 3-pulse echo inversion recovery sequence. $\trep$ is the repetition time of a sequence for all delay values. (b) Echo recovery (symbol: measurement, line: double exponential fit) after an inversion pulse acquired at $t_\text{rep} = 30~s$. (c) Inverse of relaxation time constants versus cryostat temperature performed at $t_\text{rep} = 9s$, symbols: measured, Solid line: Theory. (d) Angular dependence of the relaxation rate at a $t_\text{rep} = 20~$s. Solid line is calculated using Eq.~\ref{eq: Rff}. (e)  Inverse of relaxation time constants versus  the repetition time of the inversion recovery pulse sequence. Dashed line is a guide to the eye. All measurements are performed at $\theta =0$ except in panel (d). }
    \label{supplfig:invT1}
\end{figure}

We measure spectral/spatial diffusion limited $T_1$ with a conventional echo inversion recovery sequence. A typical curve acquired at a repetition time of $\trep = 30$s ($\gg$ spin-bath relaxation time, see the main text) is plotted in Fig.~\ref{supplfig:invT1}(b). The data fits well to a double-exponential of amplitude weights $67\%$ and $33\%$ with corresponding time constants of $T_1^a = 2.4 \pm 0.3$~ms and $T_1^b = 33 \pm 3$~ms. Interestingly, the two time constants show similar behavior across different measurement conditions: temperature, angle and repetition time [Fig.~\ref{supplfig:invT1}(c-e)]. In particular, we observe more than an order of magnitude increase in relaxation rates with increase in temperature and their saturation at 200~mK [Fig.~\ref{supplfig:invT1}(c)]. These observations can be explained using energy exchange via spin flip-flops~\cite{bottger_optical_2006, car_optical_2019, rancic_electron-spin_2022}, whose average rate over the spin distribution is given as  
\begin{align}\label{eq: Rff}
R_\text{ff} = \alpha_\text{ff}\frac{c^2 \Xi}{\Gamma_\text{inh}} \left [ \text{sech} \left(\frac{\hbar \omega_s}{2 k_\text{B} T} \right) \right ]^2,
\end{align}
where $\Xi \approx g_{ab}^4/20$ is angular averaging of dipole interactions~\cite{car_optical_2019}, $c$ the concentration of spins and $\alpha_\text{ff}$ is a constant. The corresponding calculated line in Fig.~\ref{supplfig:invT1}(c) confirms the spin flip-flop mediated $T_1$. From these, we are furthermore able to deduce the lowest spin temperature of $\sim 50$~mK, which is consistent with the one extracted from the $\gens$ measurements (see the main text).

Spin linewidth dependence of spin flop-flop rates is measured via angular dependence studies near the symmetry c-axis, which again has a good qualitative match with Eq. \ref{eq: Rff}. Indeed, a larger linewidth results in spectral dilution of the spins, thus increasing the physical separation of identical spins in the crystal~\cite{alexander_coherent_2022}.  

Extremely strong spin flip-flop rates ($>100$ times that of the spin lattice relaxation) in our heavily doped crystal and low mode-volume detection of the echo inversion recovery sequences preclude detection of a slower spin to bath relaxation time. To this end, we perform the echo inversion recovery sequences at several repetition times. A repetition time shorter than the spin to bath relaxation time results in an accumulation of spin excitation/depolarization leading to an effective higher spin temperature and thus an enhanced flip-flop rate (Eq.\ref{eq: Rff}). The extracted time constants plotted in Fig.~\ref{supplfig:invT1}(e) exactly show this behavior and remain unchanged after a $\trep > 5~$s. From these we are able to estimate the spin to bath relaxation time $\sim 1.5$~s at $\theta = 0$ ($B_0 = 254$~mT).

\section{Spin-lattice relaxation}

\begin{figure}[t!]
    \centering
\includegraphics[width=\columnwidth]{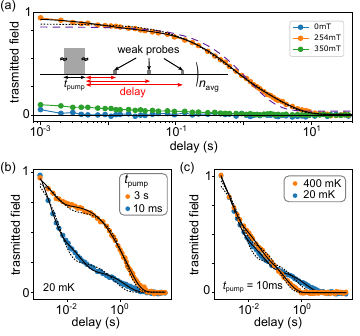} 
\caption{Measured signal decay curves across different experimental conditions using the pulse sequence shown in the inset. (a) Comparison of transmitted field at three magnetic field values at $\tpump = 3$~s. The data at $B_0=254$~mT is for the case when $I=0$ spins are resonant, and is acquired for the case with $\ginh/2\pi = 28~$MHz. Traces have been averaged 40 times for data acquired at 254~mT and 350~mT, and 6 times at 0~mT. (b) Transmitted field at different pump durations at 20~mK, and (c) different temperatures at $\tpump=10~$ms. Solid, dotted and dashed curves represent triple-, double-, and single-exponential fits.}
    \label{supplfig:pb1}
\end{figure}

\begin{figure}[t!]
    \centering
\includegraphics[width=\columnwidth]{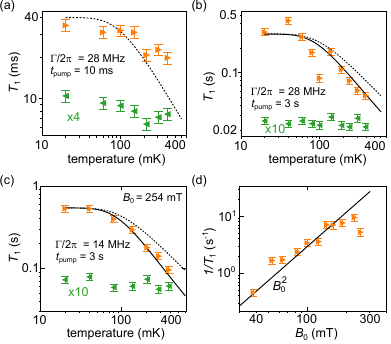} 
\caption{Behavior of additional time constants ($T^\text{f}_1$: left triangles, $T^\text{m}_1$: right triangles) across measurement conditions. (a-c) Temperature dependence at $B_0 = 254~$mT for different cases mentioned in the figure panels. Dotted line and solid line represent direct-phonon and phonon-bottlenecked processes respectively. Fast time constants have been scaled for better comparison. (d) Magnetic field dependence of $1/T^\text{m}_1$ at 20~mK. Solid line represents $B_0^2$. }
    \label{supplfig:pb2}
\end{figure}

As also mentioned in the main text, the decay of the transmitted field across the feedline following the strong saturation pump (power $\sim$ nW) shows a complex shape, and best fits to at least three exponential constants labeled as $T_{1}^\text{f},~ T_{1}^\text{m},~ \Tsl$ (fast, medium, slow). However, across all measurement conditions, we do not find any correlation between time constants and their respective weights. Several traces across varying measurement conditions are plotted in Fig.~\ref{supplfig:pb1} showing this behavior. To make sure that signals indeed arise from spins under study, the pump power is carefully chosen to avoid unwanted signals from 1) non-linearity of the resonator due to kinetic inductance of NbN film leading to a frequency shift, and 2) power dependence of the resonator's quality factor due to its coupling to TLSs present in the substrate. We furthermore compare the response of the transmitted field for the largest pump duration of 3s at several magnetic fields. While we observe a negligible response at zero magnetic field, a small response (with an amplitude less than $10\%$ at the shortest delay compared to that on the resonance data at $B_0=254$~mT) can be seen in the off-resonant case of 350~mT [see Fig.~\ref{supplfig:pb1}(a)]. Unfortunately, we were not able to perform pump-power dependence of measurements due to limitations of the setup and low signal to noise ratios at lower powers.

In any case, the small amplitude and rapid suppression of the off-resonant or background signals are less likely to affect the extraction of the slowest time constant $\Tsl$. In the following, we discuss further checks to understand the origin of the other two time constants. The temperature dependence of the slowest time constant $\Tsl$ is already discussed in the main text and therefore not repeated here. The remaining two time constants are plotted in Fig.~\ref{supplfig:pb2} against different pump duration and $\ginh$ values. We first note that the fast time constant $T_{1}^\text{f}$ (green left triangles) shows no temperature dependence under any of the probed conditions (in all panels), suggesting its origins are not linked to the erbium electron spins under study. The medium time constants $T_{1}^\text{m}$ (orange right triangles) also show negligible dependence [Fig.~\ref{supplfig:pb2}(a)] for the shortest pump duration $\tpump = 10~$ms, however, more complexity is expected at shorter delays in this case due to $\tpump$ being comparable to spectral diffusion $T_1$. The response of $T_{1}^\text{m}$, however, becomes similar to $\Tsl$ for large pump durations [Fig.~\ref{supplfig:pb2}(b,c)], as confirmed by rescaled calculations of phonon-bottlenecked spin relaxation times (solid lines) while keeping the same ratio with direct phonon relaxation times used for fitting $\Tsl$ values. The magnetic field response of $T_{1}^\text{m}$ is plotted in Fig.~\ref{supplfig:pb2}(d), and also exhibits a $B_0^2$ dependence. However, the solid line plotted in Fig.~\ref{supplfig:pb2}(d) represents values that are on average 40~times shorter compared to $\Tsl$. The latter being the actual direct-phonon spin relaxation time, consistent with previous studies. The similarity of $T_{1}^\text{m}$ with $T_{1}^\text{sl}$ indicates their source to be the same, with a plausible scenario of a phonon-avalanche mediated fast relaxation before the phonon-bottlenecked slow relaxation~\cite{brya_paramagnetic_1965}. Though more measurements are needed to ascertain the origin, as other possible effects such as radiation damping of spin polarization can also produce similar signal response~\cite{schweiger_principles_2001}.

\bibliography{Er_Relaxation}

\end{document}